\documentclass[aps,pre,epsfigure,onecolumn,superscriptaddress,notitlepage]{revtex4-1}
\usepackage{dcolumn} 
\usepackage{bm} 
\usepackage{graphicx}
\usepackage{amsmath}    
\usepackage{latexsym}
\usepackage{amsfonts}   
\usepackage{amssymb}
\usepackage{array}      
\usepackage{epsfig}
\usepackage{txfonts}
\usepackage{color}
\usepackage[colorlinks=true,linkcolor=blue,urlcolor=blue,citecolor=blue]{hyperref}
\usepackage{hyperref}
\usepackage{enumerate}
\usepackage{bbold}
\usepackage{epstopdf}
\usepackage{cleveref}
\usepackage{appendix}
\usepackage{float}
\newcommand{\ntu}{Division of Physics and Applied Physics, School of Physical 
	and Mathematical Sciences, Nanyang Technological University, Singapore}
\newcommand{\cnr}{CNR--SPIN, Dipartimento di Scienze Fisiche, Universit\`a di 
	Napoli Federico II, I-80126, Napoli, Italy}
\newcommand{\ihpc}{Institute of High Performance Computing, Agency for Science Technology and Research, Singapore}

\begin{document}
\title{Designing Phononic Band Gaps with Sticky Potentials}	
\author{Yuanjian Zheng} 	
\affiliation{\ntu}
\author{Shivam Mahajan} 
\affiliation{\ntu}
\author{Joyjit Chattoraj} 
\affiliation{\ihpc}
\author{Massimo Pica Ciamarra} 
\affiliation{\ntu}
\affiliation{\cnr}
\begin{abstract} 
Spectral gaps in the vibrational modes of disordered solids are key design elements in the synthesis and control of phononic metamaterials that exhibit a plethora of novel elastic and mechanical properties. However, reliably producing these gaps often require a high degree of network specificity through complex control optimization procedures. In this work, we present as an additional tool to the existing repertoire, a numerical scheme that rapidly generates sizeable spectral gaps in absence of any fine tuning of the network structure or elastic parameters. 
These gaps occur even in disordered polydisperse systems consisting of relatively few particles ($N \sim 10^2-10^3$).
Our proposed procedure exploits sticky potentials that have recently been shown to suppress the formation of soft modes, thus effectively recovering the linear elastic regime where band structures appear, at much shorter length scales than in conventional models of disordered solids. 
Our approach is relevant to design and realization of gapped spectra in a variety of physical setups ranging from colloidal suspensions to 3D-printed elastic networks.	   
\end{abstract}	
\maketitle
\section{Introduction}	
Low energy excitations of vibrational modes in disordered solids have been a subject of recent interest, for it presents an exciting approach to the construction of tunable metamaterials with a wide range of applications. From its collective contributions to bulk acoustic or sound attenuation properties \cite{CummerAlu2016}, to the remarkable precision in conformational changes made available with the advent of directed mechanical response \cite{BertoldiVanHecke2017,ShimReis2012,NicolaouMottler2012}, these soft modes have offered an unprecedented level of control over the design of modern metamaterials due to its remarkable localization and topological properties \cite{RocklinMao2017,SouslovVitelli2017,SusstrunkandHuber2016,SigmundandJensen2003}. In particular, recent phononic applications originating from these concepts include the design of auxetic materials with negative possion ratio \cite{ReidDePablo2018, ReidDePablo2019, ZhaoFang2019}, allosteric inspired nano- or macro-scale levers capable of eliciting a targeted mechanical response at a distance \cite{RocksNagel2017,RocksKatifori2019}, and the experimental realization of materials with negative effective stiffness \cite{Lakes1987,LakesWang2001, BrunetMondainMonval2014}. 

To robustly implement these novel features in a sustained and controllable manner requires the existence of spectral band gaps that isolate the specific desired functionality of low energy excitations from the remainder of its spectrum that constitute noise to the intended response. 
As such, understanding and controlling the properties of spectral gaps forms an important aspect to metamaterial design that goes beyond simply optimizing for the displacement field of the low frequency eigenmodes. 
This concern for the spectral architecture has motivated studies on the formation and tunability of band gaps in disordered photonic systems, that have recently also been incorporating hyperuniform point distributions that display peculiar behavior in Fourier space and their resulting structure factor \cite{ManChaikin2013, FlorescuSteinhardt2009, LiBi2018}. 
Previous studies addressing phononic or acoustic band gaps in discrete elastic systems have largely focused on ordered or quasi-periodic systems \cite{MousanezhadVaziri2015, MitchellIrvine2018, NashIrvine2015, WangBertoldi2015}, with far less attention paid to spectral gaps of disordered solids \cite{RonellenfitschDunkel2019}. 

In this work, we present a numerical scheme capable of rapidly generating phononic spectral band gaps that dictate the robustness of programmable mechanical behavior in disordered elastic networks. 
Unlike existing strategies that depend on optimizing linear response \cite{RonellenfitschDunkel2019} or some secondary cost functions \cite{RocksNagel2017,RocksKatifori2019}, this scheme robustly generates spectral gaps with no fine tuning of the elastic and topological properties. 
As such, our approach serves not only as a standalone algorithm, but can also be used to generate precursors used in combination with the various existing optimization techniques to accentuate particular functionalities, while having the assurance of sizeable spectral gaps at the vicinity of a pre-determined frequency. 

Phononic band gaps are naturally present in all systems as long as their size is larger than their characteristic disorder length scale. 
In this limit, the vibrational spectrum exhibits well defined phonon gaps that characterizes linear elasticity. 
However, for conventional systems of particles interacting via purely repulsive \cite{WeeksAndersen1971} or Lennard-Jones(LJ)-like \cite{KobAndersen1995} potentials, this disorder length scale is large. 
For instance, in a study of two-dimensional LJ systems, convergence to the elastic limit was only observed for systems comprising millions of particles \cite{TanguyBarrat2002}. 
From a design perspective, devising disordered and gapped mass-spring networks of a small manageable size that one might conceivably construct for example via 3d printing techniques, thus becomes the key issue of concern. 
Here, we introduce an approach to do so. 

This approach in essence involves first obtaining energy minimal configurations of pairwise interacting particles mediated through a family of sticky potentials that are short range repulsive, but attractive at intermediate distances of separation \cite{ChattorajCiamarra2020,WangDeng2020}. 
Recent work suggest that such interactions result in the drastic reduction in the availability of soft vibrational modes in its jammed state \cite{LopezLerner2020a, LopezLerner2020b}, and thus these networks exhibits gapped spectra reminiscent to that of linear elasticity at much shorter lengths scale compared to their WCA or LJ-like counterparts, despite being inherently disordered from a structural perspective \cite{LopezLerner2020a}. 
The arrived energy minimal configuration can then be mapped onto a harmonically interacting linear elastic network that serves as the blue print for construction of the target phononic metamaterial. 

In the following, we first introduce in Sec. (\ref{section:sticky_potentials}) the general class of sticky potentials and energy minimization protocols specific to this work, before briefly discussing elastic properties of the networks derived from energy minimal configurations. 
In Sec. (\ref{sec:spectral}), we quantify the occurrence statistics of the spectral gap and their corresponding frequencies, showing that they remain robust and highly predictable over a range of (\ref{sec:rho}) densities and (\ref{sec:r_c}) parameters of the potential. 
Moreover, we show that these spectral gaps remain pronounced even in small systems (\ref{sec:size}), which are often of particular interest to practical realizations. 
In Sec. (\ref{sec:pre_stress}), we demonstrate how removal of the pre-stress contribution can further enhance performance of the spectral gaps. 
Lastly we conclude our work by summarizing in a schematic (\ref{sec:conclusion}), the step-wise procedures involved in obtaining robust gapped spectra in linear elastic networks.

\section{Sticky potentials and elastic networks}
\label{section:sticky_potentials}
The key towards reliable generation of phononic band gaps in small disordered systems lies in the emergent collective properties of particles interacting via a class of ``sticky" potentials \cite{ChattorajCiamarra2020}. 
In this section, we review the functional form of this potential that has been adapted specifically for this work, and briefly discuss its various properties. 

Traditional atomistic models of disordered solids have largely focused on the purely repulsive Weeks-Chandler-Andersen (WCA) \cite{WeeksAndersen1971} potential, or on the Lennard-Jones (LJ) one \cite{KobAndersen1995} which possesses a slowly decaying attractive tail $ \propto 1/r^6$, that captures qualitative behavior of physical Van-der-Waals interactions. 
Sticky potentials seek to vary this effective range of the attractive tail at intermediate length scales. 
This seemingly modest modification to the local potential has far reaching consequences to the dynamical behavior of its relaxation dynamics \cite{ChattorajCiamarra2020} and the elastic properties \cite{KoezeTIghe2020} of its jammed configuration at sufficiently high densities \cite{WangDeng2020} and is also known to result in systems with negative thermal expansion \cite{RechtsmanTorquato2007}. 

More importantly, the energy minimal state in solids or otherwise inherent state \cite{StillingerWeber1982}, where pressure $(p)$ and shear modulus $(\mu)$ remains positive definite, also exhibits phononic spectra $\{\omega_n\}$ and density of states $\mathcal{D}(\omega)$ that differ significantly from the typical behavior of jammed or LJ-like disordered solids \cite{LopezLerner2020a,LopezLerner2020b}. 
As observed in all disordered solids
\cite{Mizuno2014,LernerBouchbinder2016,Kapteijns2018,Richard2020,Rainone2020}, the vibrational density of states of these systems is the superposition of a 
quartic scaling behavior regime $\mathcal{D}(\omega)\sim A_4\omega^4$ comprising quasi-localized modes, and of a Debye-like contribution, $\mathcal{D}(\omega)\sim A_4\omega^{d-1}$ in $d$ spatial dimensions.
However, quasi-localized soft modes in sticky potentials are severely suppressed \cite{LopezLerner2020a,LopezLerner2020b}, e.g. the parameter $A_4$ is unusually small. 
This work thus focuses on how one might exploit the natural suppression of these quasi-localized low frequency modes to produce robust gapped spectra that adheres closely to linear elasticity even in small disordered systems. 

We note that several variants of these sticky potentials exists in the current literature \cite{ChattorajCiamarra2020, LopezLerner2020a, LopezLerner2020b,WangDeng2020,KoezeTIghe2020}, with specifics of its parameterization playing a non-trivial role in the behavior of physical observables \cite{LopezLerner2020a, LopezLerner2020b}. These variabilities seem preliminarily related to the degree of ``stickiness", an emergent measure of the relative attractive to repulsive forces present in the energy minimal state \cite{LopezLerner2020a}. Here, instead of discussing the various trade-offs involved, we focus on a particular version \cite{ChattorajCiamarra2020} of sticky potentials,
\begin{align}
\frac{U(r_{ij})}{\varepsilon}=
\begin{cases}
4 \left[\left( \frac{\sigma_{ij}}{r_{ij}}\right)^{12} - \left( \frac{\sigma_{ij}}{r_{ij}}\right)^6\right] & \text{if } r_{ij} \leq r^{min}_{ij} \equiv 2^{1/6} \sigma_{ij}
\\
a_0\left( \frac{\sigma_{ij}}{r_{ij}}\right)^{12} - a_1 \left( \frac{\sigma_{ij}}{r_{ij}}\right)^6 +\sum^{3}_{l=0} c_{2l} \left(\frac{\sigma_{ij}}{r_{ij}}\right)^{2l} & \text{if } r^{min}_{ij} \leq r_{ij} \leq r_c
\label{eq:potential}
\end{cases}
\end{align}
and suggest corresponding parameter values that reliably produce gapped spectra with high fidelity. 

	\begin{figure}[htpb!]
	\includegraphics[width=0.98\columnwidth]{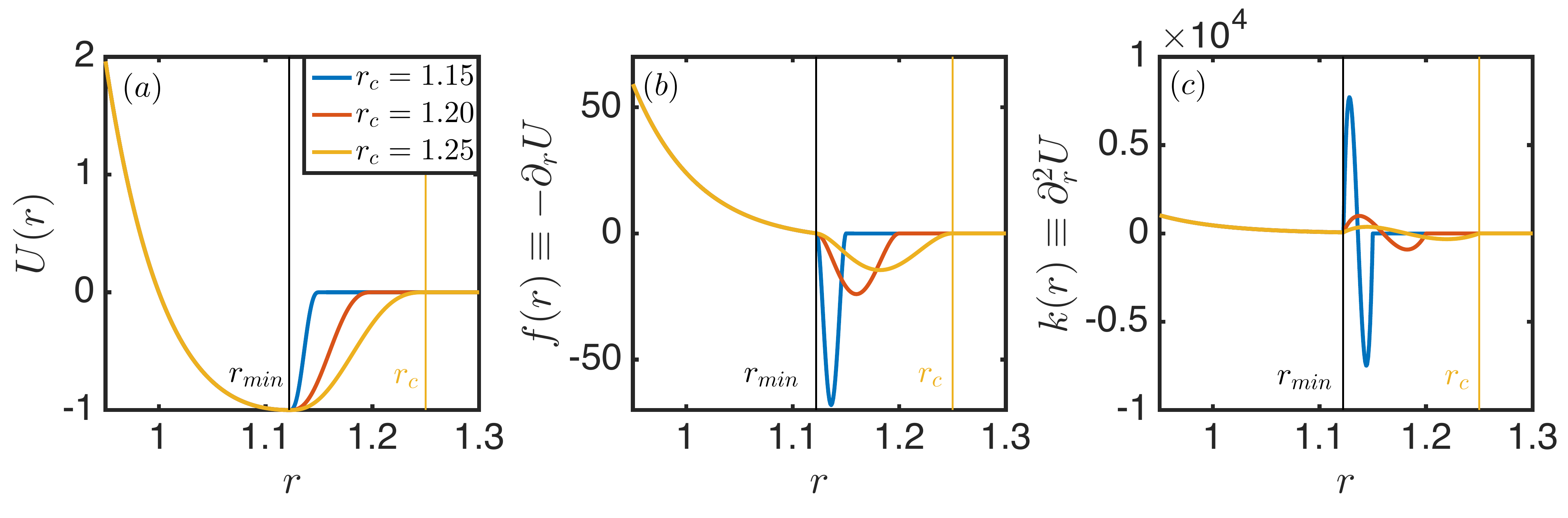}		
	\caption{(a) Pairwise potential- $U(r)$, (b) bond tensions - $f(r)$ and (c) effective stiffness - $k(r)$ for various parameter values $r_c$ of the interaction potential {eq.(\ref{eq:potential}}) considered. $\langle \sigma_i \rangle$ and $\varepsilon$ are the units of length and energy respectively. Note the broadening domain of negative stiffness in $r$ with increasing $r_c$. 
	}
	\label{fig:potential}
\end{figure}

In this potential (\ref{eq:potential}), particles at pair separations $r_{ij} \leq r^{min}_{ij} \equiv 2^{1/6} \sigma_{ij}$ interact repulsively via Kob-Andersen / Lennard-Jones-like interactions (LJ) \cite{KobAndersen1995} similar to purely repulsive WCA potentials \cite{WeeksAndersen1971} where $\sigma_{ij} \equiv (\sigma_i + \sigma_j)/2$ is the effective length scale that characterizes polydispersity $\sigma_i$ of interacting particles. In addition to this contact repulsion at extremely short length scales, particles also attract each other at intermediate distances $r^{min}_{ij}\leq r_{ij} \leq r_c$, in which $r_c$ is a cut-off parameter where the potential vanishes. The remaining coefficients $a_0, a_1$ and $c_{2l}$ are then uniquely obtained by solving equations that ensure continuity in $U(r)$ and its first $\partial_r U(r)$ and second derivatives $\partial^2_r U(r)$ at distances $r^{min}_{ij}$ and $r_c$.
	
Such a parameterization allows for the repulsive ($f(r) > 0$) part of the interaction to be unchanged for different values of $r_c$ while varying the domain in $r_{ij}$ where interactions are attractive ($f(r) < 0$) (see Fig. \ref{fig:potential}a). In the limit of $r_c \to r^{min}$, $U(r)$ becomes discontinuous at $r_{min}$ such that non-zero region of $U(r)$ is purely repulsive and resembles qualitatively the WCA potential, while $r_c \to \infty$ recovers the long range interacting LJ potential. 
Previous numerical studies indicate that $r_c \geq 2.4$ essentially recovers the LJ limit in which the phononic spectra is known to be continuous \cite{ChattorajCiamarra2020}. 
As such, interesting parameter regimes of the sticky potentials are restricted to intermediate values of $r_c \sim 1.2$. 

Unlike purely repulsive systems, the resulting bond tensions $f\equiv -\partial_r U$ and stiffnesses $k \equiv \partial^2_r U$ between particles derived from sticky potentials can assume negative values as seen in Fig. \ref{fig:potential}(b) and (c) respectively. 
We note that while $r_c \to \infty$ reduces the magnitude of extreme values in the local bond stiffnesses which may in principle ease its construction, the broadening of regime in $r_{ij}$ where $k$ is negative as $r_c$ increases, renders such systems to be impractical especially for the purpose of constructing elastic networks, and thus may instead be detrimental to its design. 
Moreover, a relatively large $r_c$ as compared to the average inter-particle distance $\rho^{-1/d}$ results in extended interactions with other particles beyond its immediate neighbors. This further complicates the resulting elastic network and also increases the number of linear springs involved. These considerations and their effects on the vibrational spectra are deliberated in greater detail in the subsequent sections.

\begin{figure}[hbtp!]
	\includegraphics[width=0.75\columnwidth]{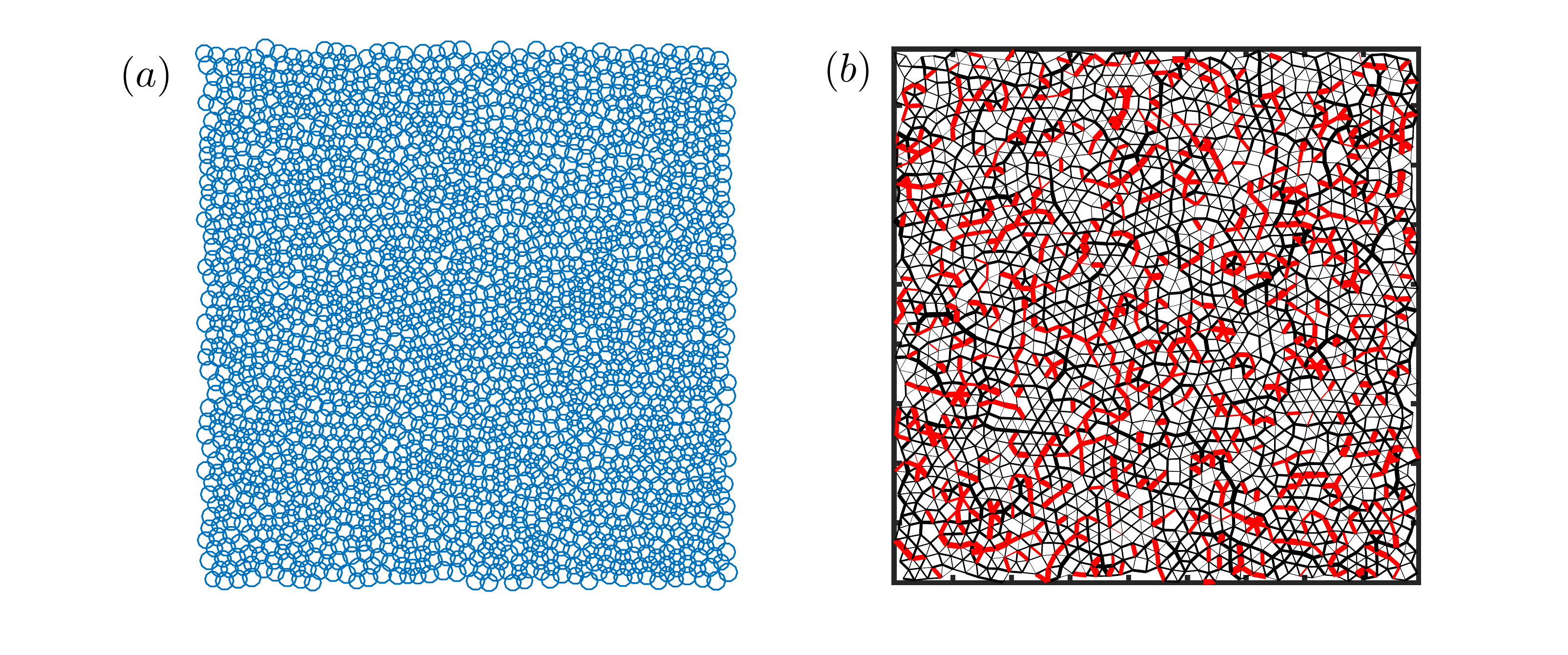}		
	\caption{(a) Energy minimal configuration of $N=2000$ polydisperse particles at density $\rho=1$ and $r_c=1.2$ . (b) Corresponding elastic network - $\left\{ k_{ij}, l_{0,ij} \right\}$ consisting of attractive (red) and repulsive (black) interactions derived from the disordered configuration. Thickness of each line are scaled to represent the effective pairwise stiffness $k_{ij}$ of each bond.
	}
	\label{fig:network_representation}
\end{figure}

For a given $r_c$ and number density $\rho$ , the sticky potential can be used to deterministically generate particle configurations in arbitrary dimensions using a variety of energy minimization algorithms that ranges from gradient descent to monte carlo based methods \cite{numerical_methods_a,numerical_methods_b}. In this work, we implement a conjugate gradient algorithm \cite{Fletcher1976} to obtain particle configurations starting from initial states generated from a homogeneous Poisson point process, with all data presented unless otherwise stated, derived from systems of $N=2000$ particles in two dimensions. $\sigma_{i}$ selected from a uniform distribution within $[0.8, 1.2]$ such that the length scale of the system is given by $\langle \sigma_i \rangle=1$ as conventionally adopted in previous studies of sticky potentials \cite{ChattorajCiamarra2020}.
The system is highly polydisperse, the standard deviation of the $\sigma$ distribution being $\simeq 12\%\sigma$, and hence lack both translational and positional order.
$\varepsilon \equiv 1$ sets the energy scale of the system, and all particles have mass $m=1$.

Termination of the minimization is imposed upon mechanical equilibrium based on a threshold defined by magnitude of the largest net force acting on each particle - $ \max_{i} \vert \vert \sum_{j \neq i} \vec{f}_{ij} \vert \vert \leq 10^{-12}$. 

The linear response regime of this energy minimal configuration is investigated by Taylor expanding to second order the energy of the system around the equilibrium configuration~\cite{Ashcroft76}.
In doing this operation, one maps the positions of particles in the energy minimal configurations $\{\vec{R}_{i}\}$ as shown in Fig. \ref{fig:network_representation}(a) on to an equivalent elastic network Fig.(\ref{fig:network_representation}(b) where $\{\vec{R}_{i}\}$ now represent the position of nodes, and pairwise interactions ($r_{ij} \leq r_c$) are replaced by effective harmonic springs of stiffness $k_{ij}=\partial_r U(r_{ij})$ and rest lengths $l_{0,ij}=r_{ij}+(f_{ij}/k_{ij})$. 
The response of this effective elastic network informs on vibrational properties of the energy minimal configuration in the linear response regime, and thus in turn fully determine the phononic spectrum of the material. 
In Fig. \ref{eq:elastic}, we show the statistical features of elastic properties of derived networks for an ensemble of $N_{samples}=2500$ configurations for various $\rho$ evaluated at  $r_c=1.2$. 

\begin{figure}[htpb!]
	\includegraphics[width=0.75\columnwidth]{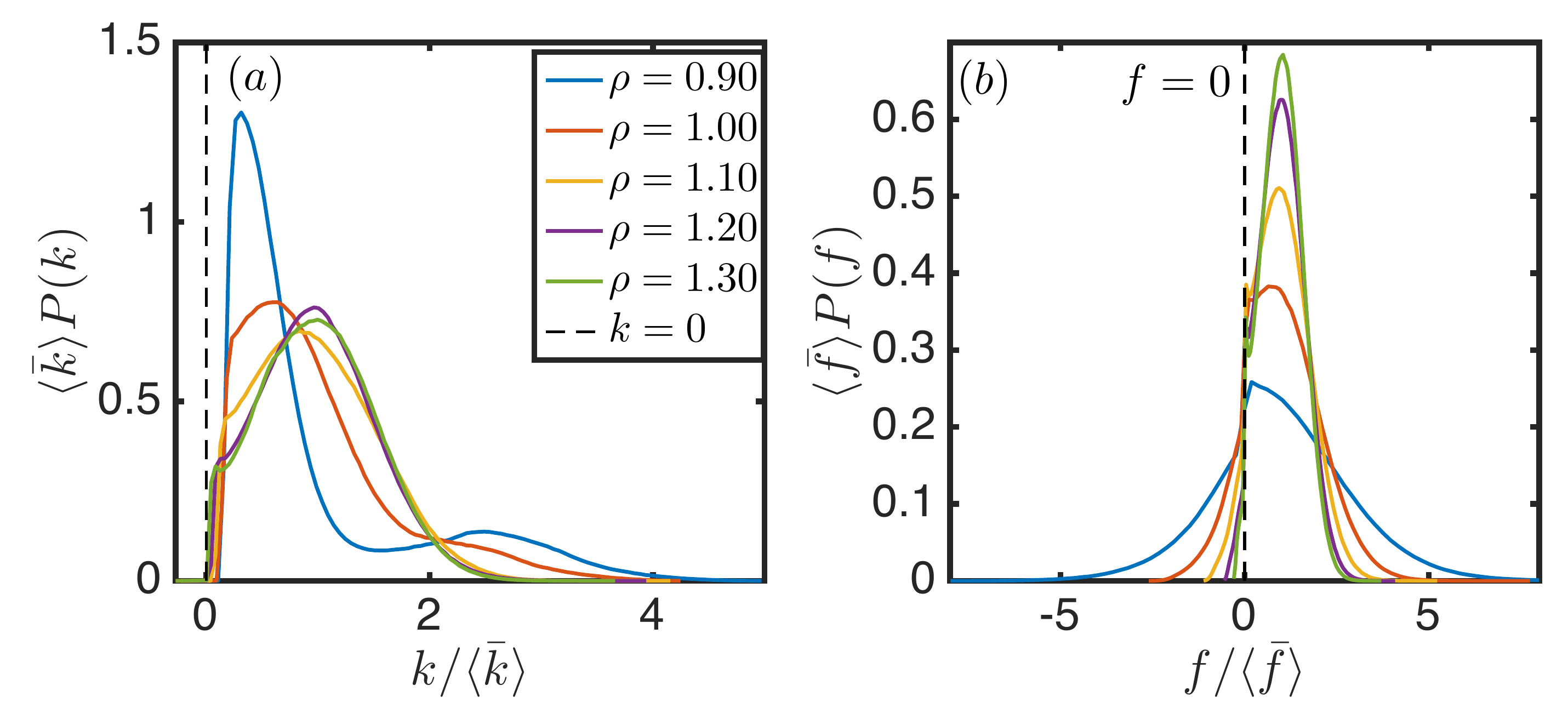}		
	\caption{Probability density function of bond (a) stiffness - $\bar{k}$ and (b) tensions - $f$ for $r_c=1.2$ at various number densities $\rho$. Unlike purely repulsive potentials, pairwise interactions can be attractive in nature, and bonds can in principal assume negative stiffness. However, for suitably small values of $r_c$ (given $r_c > r^{min}_{ij}$), $k_{ij} < 0$ occurs with much lower probability for a wide range of densities.}
	\label{fig:elastic_properties}
\end{figure}

While potentials generically permit the existence of negative stiffness $k < 0$, a good choice of $r_c$ that restricts the domain at which this happens can significantly reduce the probability of encountering unconducive environments for a wide range of densities (Fig.\ref{fig:elastic_properties}(a)). 
Indeed, $k < 0$ is undesired for material design, as it is difficult to fabricate unit springs or linear elastic segments with negative stiffness. 
Moreover, we note that the distribution of $k$ can exhibit bi-modalities at sufficiently high $\rho$ where the average inter-particle length scale enters deep into the repulsive regime and is thus related to the existence of contacts with particles in the first and second coordination shell as mentioned above. On the other hand, the distribution of bond tensions (Fig.\ref{fig:elastic_properties}b) remains uni-modal at all densities considered but attractive interactions ($f < 0$) diminish with increasing $\rho$, indicating the reduction of stickiness driven by a change in density that was also observed in \cite{LopezLerner2020a}. These properties collectively may provide additional avenues of tunability for multi-functionalities that depend on local stiffnesses of the material \cite{RocksKatifori2019} that are no doubt interesting possibilities that lies beyond the scope of this work . 

\section{Spectral properties}
Having obtained the elastic networks representing energy minimal configurations of sticky potentials in (\ref{section:sticky_potentials}), we evaluate and discuss the novel spectral properties of sticky solids in this section. In general, the vibrational modes represent the elementary excitations of a (meta)stable configuration and can be obtained from the dynamical matrix or otherwise hessian of its potential:
\begin{equation}
\mathcal{M}^{\alpha \beta}_{ij}\equiv \frac{\partial^2 U}{\partial R_i^\alpha \partial R_j^\beta}
\label{eq:hessian}
\end{equation}
where $\alpha,\beta$ labels the indices of orthogonal axes and $i,j$ indexes the particles. Upon diagonalization, the hessian yields normal modes $\vert n \rangle$ with eigenfrequencies $ \omega_n =\sqrt{-\lambda_n} $ where $\lambda_n$ represents eigenvalues of the $n$-th mode.

\begin{figure}[htpb!]
	\includegraphics[width=0.62\columnwidth]{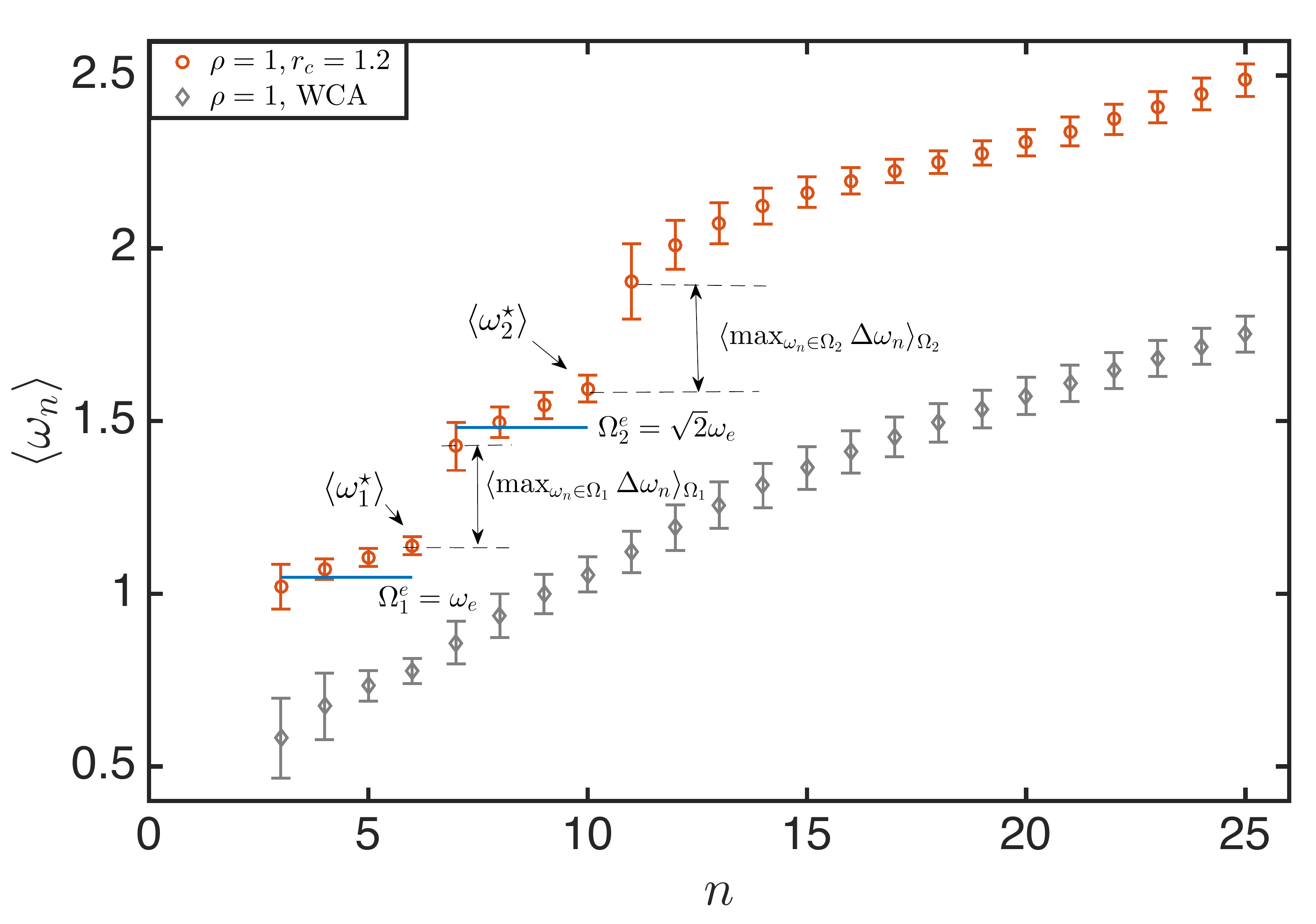}		
	\caption{Average vibrational frequencies $\langle \omega_n \rangle$ (standard deviation represented by the error bars) for each mode number $n$ over $N_{samples}=2500$ configurations of $N=2000$ particles each. Blue horizontal lines indicate the frequency of normal modes as predicted by linear elasticity eq. (\ref{eq:elastic}). Unlike systems mediated by purely repulsive interactions such as WCA (gray), energy minimal states of disordered systems driven by sticky potentials eq. (\ref{eq:potential}) acquire gapped spectral behavior reminiscent of linear elasticity even in small systems. }
	\label{fig:wca}
\end{figure}

For a typical disordered system interacting via repulsive WCA potential, the length scale in two dimensions and corresponding system sizes at which continuum elasticity sets in is tremendously large ($\geq 4 \times 10^4$) \cite{TanguyBarrat2002} such that phononic excitations are very low in frequency and thus do not appear in significantly smaller systems ($N \sim10^2-10^3$). As such, the low frequency branch is dominated by quasi-localized soft modes which results in a continuous spectra in which no spectral gaps exists as we show in Fig.\ref{fig:wca}, by plotting the average frequency $\langle \omega_n \rangle$ of a given mode number $n$ calculated for systems of size $N=2000$.
\label{sec:spectral}

Now, unlike their WCA or LJ counterparts, solids of sufficient stickiness suppresses the density of quasi-localized soft modes \cite{LopezLerner2020a} such that the spectrum is not continuous and gapped as seen in Fig. \ref{fig:wca}. More importantly, the typical occurrence of these gaps and frequencies of the quasi-degenerate branches in $\{\omega_n\}$, all closely resemble predictions made based on the fundamental frequency of the transverse mode $\omega_e$ suggested by linear elasticity, 
\begin{equation}
\omega_e=\sqrt{\frac{2\pi}{L}\left(\frac{\mu}{\rho}\right)^{1/2}}
\label{eq:elastic}
\end{equation}
which we evaluate and plot as blue horizontal lines in Fig. \ref{fig:wca} for non-trivial modes in the first two branches of the spectrum. 
From a design perspective, the question of interest is how to robustly generate and maximize the observed gaps.

To investigate this issue and formally characterize properties and assess robustness of these spectral gaps, we define a dimensionless characteristic gap frequency ($\omega_i^{\star}$) for the $i$-th linear elastic branch:
\begin{equation}
\omega_i^{\star}=\max_{\omega_n \in \Omega_i} \omega_n
\label{eq:characteristic_frequency}
\end{equation}
that marks the onset (lower) frequency of the gap and a corresponding dimensionless gap parameter ($\gamma_i$) between the $i$ and $i+1$ linear elastic branch:
\begin{equation}
\gamma_i=\frac{ \max_{\omega_n \in \Omega_i} \Delta \omega_n}{ \langle \Delta \omega_{n}\rangle_{ \Omega_i}}
\label{eq:gap_parameter}
\end{equation}
where $\Delta \omega_n =  \omega_{n+1}-\omega_n$ and  $\langle \dots \rangle_{\Omega_i}$ denotes the average over modes in the $i$-th branch of the vibrational spectrum. Visual representation of these quantities are shown schematically in Fig.\ref{fig:wca}. 

In the following subsections, we examine the properties and consistency of these spectral parameters derived for a ``clean" ensemble consisting of post-selected samples that strongly adheres to linear elasticity under various conditions. 
The clean ensemble consists of samples that do not contain rattlers or bonds with negative stiffness, and also do not include samples where either of the gaps occurs at a mode number not in accordance to linear elasticity. 
This is done for the ease of consistency in physical realizations that do not require fine tuning. Specifically we investigate the $\rho$ and $r_c$ dependence of this ensemble in \ref{sec:rho} and \ref{sec:r_c} respectively, and show that these gaps can persists even in systems as small as $N=500$ in section \ref{sec:size}.

\subsection{$\rho$ dependence}

\label{sec:rho}

\begin{figure}[hbtp!]
	\includegraphics[width=0.68\columnwidth]{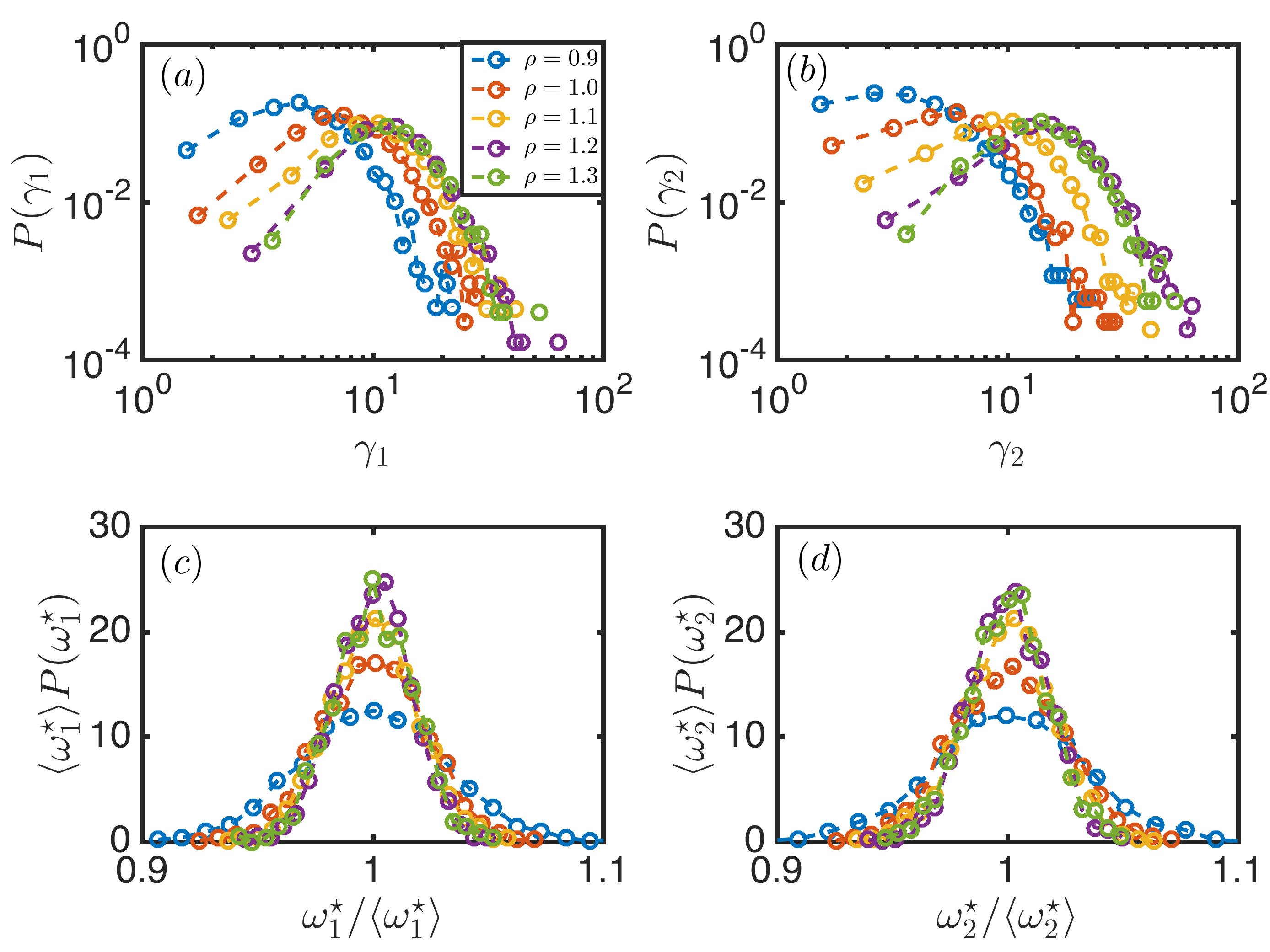}		
	\caption{Probability density function of the respective gap parameters (a) $\gamma_1$ (b) $\gamma_2$ and the characteristic gap frequency (c) $\omega_1^\star$ (d) $\omega_2^\star$ for $r_c=1.2$ at various number densities $\rho$. Note the reduced fluctuations in $\omega_i^\star$, increased values of $\gamma_i$ and saturation of the various distributions at high densities $\rho > 1.2$, which indicates the robust generation of phononic band gaps. }
	\label{fig:spectral_distribution_rho_variation}
\end{figure}

\begin{figure}[htbp!]
	\includegraphics[width=0.92\columnwidth]{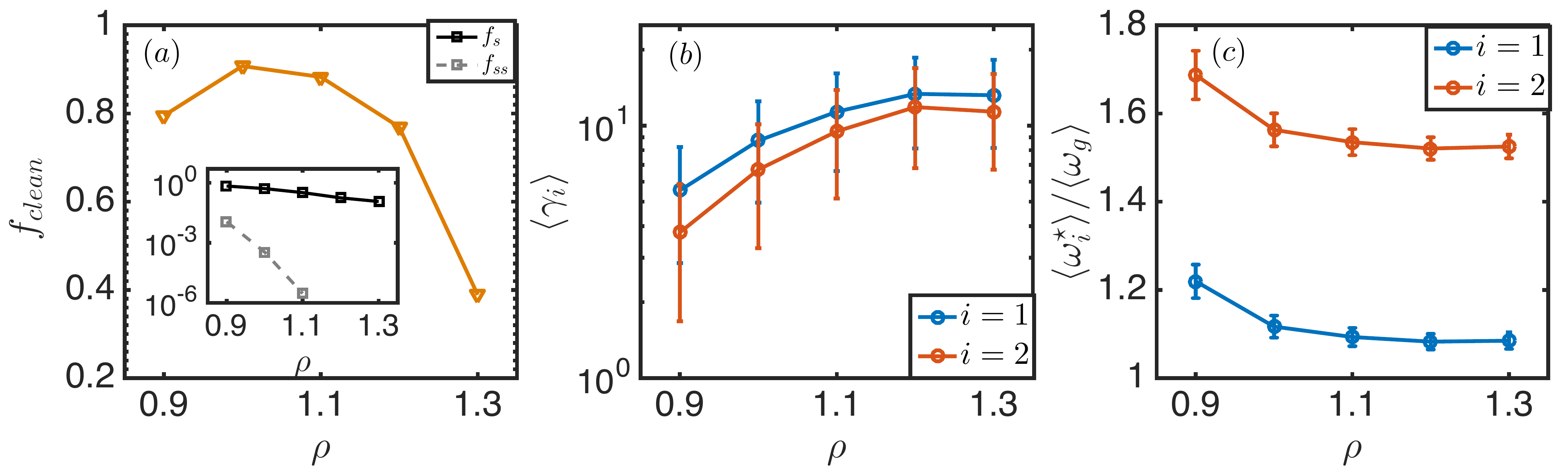}		
	\caption{(a) Fraction of samples ($f_{clean}$ - orange) that display a phononic band gap at a mode number in accordance to linear elasticity and which do not posses rattlers or bonds with negative stiffness for various values of $\rho$ at $r_c=1.2$. [Inset] fraction of sticky ($f_s$ - black) and super sticky ($f_{ss}$ - gray) particles at various densities. Note the depletion of super sticky particles and the reduction in$f_{clean}$ with increasing $\rho$. Ensemble average (b)  $\gamma_i$ and (c) $\omega^\star_i$ scaled over $\langle \omega_g\rangle$- the average lowest non-trivial frequencies, over the samples that constitute $f_{clean}$.}
	\label{fig:spectral_mean_rho_variation}
	
\end{figure}

In Fig. \ref{fig:spectral_distribution_rho_variation}(a,c) we show the distribution in $\gamma_i$ for various $\rho$ with $r_c=1.2$ for the $i=1,2$ linear elastic branches respectively. We see that with increasing density, the profiles of gap distributions comprising of clean samples shift towards higher values, indicating that the non-dimensional features of these gaps are accentuated at higher frequencies. This observation is also supported by the largely increasing trend in the average of this distribution as shown in Fig. \ref{fig:spectral_mean_rho_variation}(b). 

Furthermore, we note that the low frequency tail of $P(\gamma_i)$ decreases with increasing $\rho$ suggesting that the proportion of systems with $\gamma_i \sim \mathcal{O}(0)$ is vanishing at sufficiently high density. At the same time, the distributions of characteristic frequencies at which the gap occurs Fig. \ref{fig:spectral_distribution_rho_variation}(b,d), and their respective mean values Fig. \ref{fig:spectral_mean_rho_variation}(c), also suggest that $\omega^{\star}_i$ typically increases with $\rho$, while the fluctuations about its mean values decreases, resulting in a strongly peaking distribution in $\omega_i^\star$ for high density configurations.

For the intended purpose of robustly producing physically implementable spectral gaps in materials with high fidelity, a consistently larger $\gamma_i$ with smaller fluctuations in $\omega^{\star}_i$ is intuitively desired. This might suggest that higher parameter values of $\rho$ may be conducive to the production of gapped systems. However we show in Fig.\ref{fig:spectral_mean_rho_variation}(a) that the fraction of samples considered clean and thus constituting the ensemble corresponding to the observed distributions, is a non-monotonic function of the density. In fact, it dramatically decreases at higher densities ($f_{clean} \sim 0.4$ for $\rho=1.3$), which implies that the intuitive notion of preferring higher $\rho$ may not be entirely productive in recovering linear elasticity in small systems. Indeed, with increasing densities, the typical inter-particle length scale $ \sim \rho^{-1/d}$ shifts towards the purely repulsive regime in $U(r)$, which thus suggest behavior that rapidly approaches ungapped WCA driven systems with increasing density. 

This perspective on the role of $\rho$ in gap generation is further confirmed by studying the average fraction of sticky $f_s$ and super-sticky $f_{ss}$ particles across all samples generated, (i.e inclusive of ``unclean" configurations) which we show in the inset of Fig.\ref{fig:spectral_mean_rho_variation}(a). Sticky particles refer to nodes with at least one attractive interaction, while super sticky particles are those that only interact attractively (i.e no repulsive interaction). Note that no super sticky particles are present in our data for the two highest densities considered, indicating that $f_{ss} < 10^{-6}$ for $\rho \geq 1.2$. Now, we see that although the behavior of $\gamma_i$ and $\omega^\star_i$ is desired for higher $\rho$ within the ensemble of clean samples, the emergent stickiness monotonically decreases, resulting in the presence of fewer clean samples. 
This nuance in the $\rho$ dependence may be of concern to material design, and its relevance is dependent on whether post selection is part of the protocol in question. Densities within the range - $0.9 \lessapprox \rho \lessapprox 1.2$ are generally recommended.

\subsection{$r_c$ dependence}
\label{sec:r_c}

\begin{figure}[htbp!]
	\includegraphics[width=0.68\columnwidth]{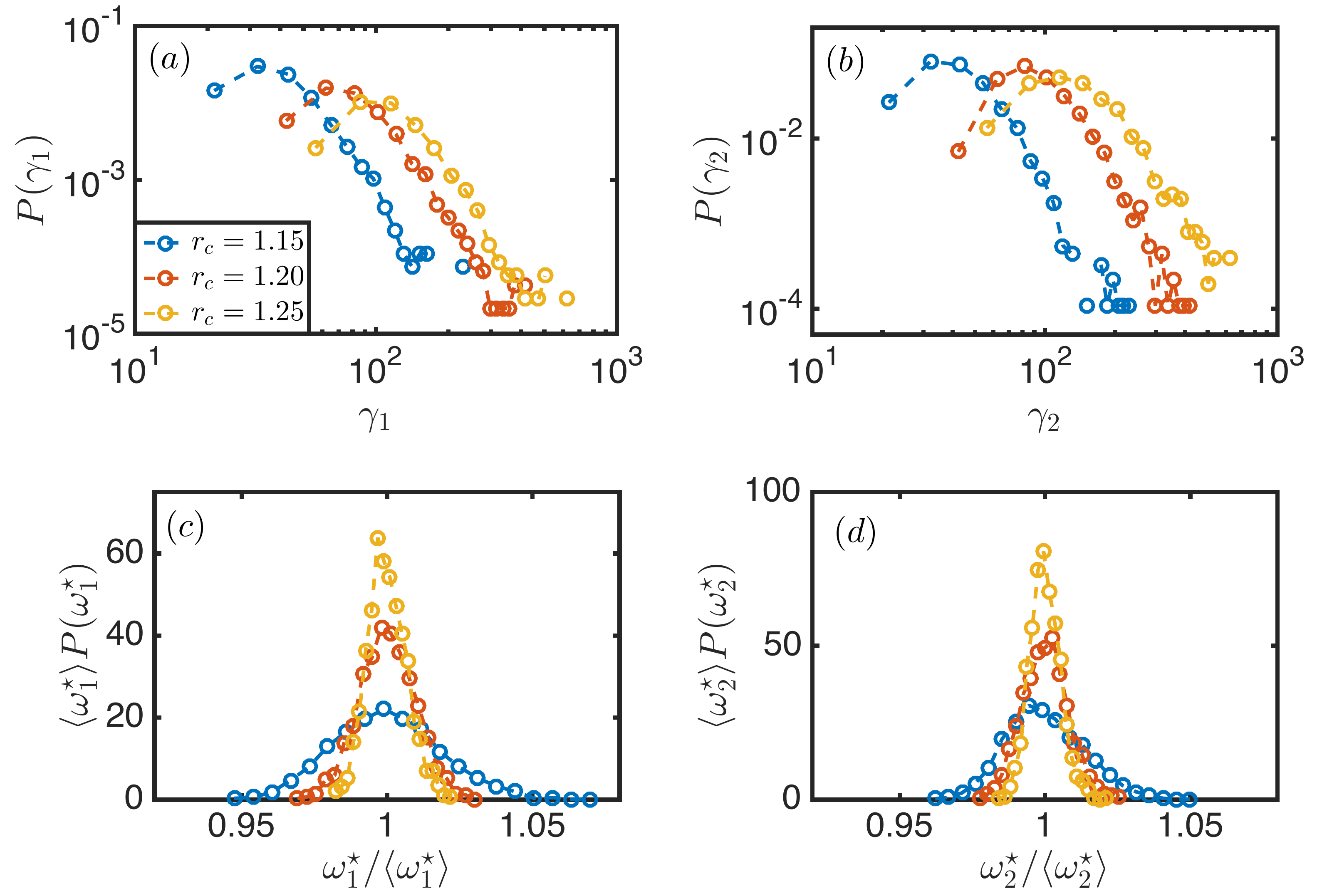}		
	\caption{Probability density function of the respective gap parameters (a) $\gamma_1$ (b) $\gamma_2$ and the characteristic gap frequency (c) $\omega_1^\star$ (d) $\omega_2^\star$ for $\rho=1$ at various values of $r_c$. Note the increase in typical values of $\gamma_i$ and decrease in fluctuations of $\omega_i^\star$ with increasing $r_c$}
	\label{fig:spectral_distribution_r_c_variation}
\end{figure}

\begin{figure}[htbp!]
	\includegraphics[width=0.92\columnwidth]{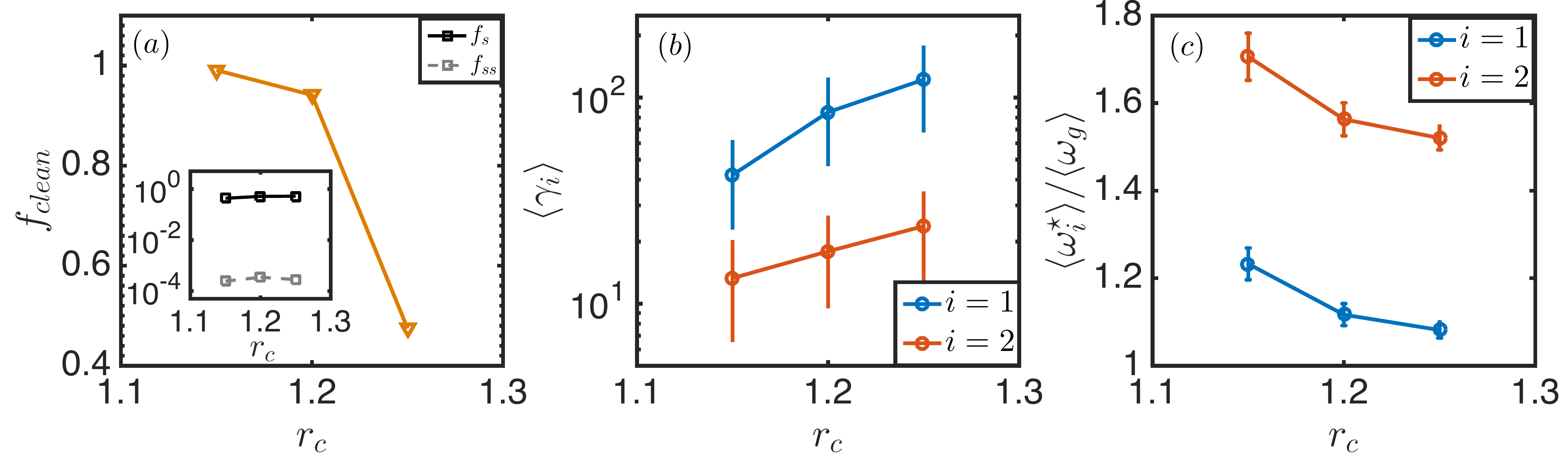}		
	\caption{(a) Fraction of samples that exhibit a spectral gap at a mode index in accordance to linear elasticity that do not contain any rattlers or bonds with negative stiffness - $f_{clean}$ (orange) for various values of $r_c$ at $\rho=1$. [Inset] fraction of sticky ($f_s$ - black) and super sticky ( $f_{ss}$ gray) particles at various $r_c$ at $\rho=1$. Average (b)  $\gamma_i$ and (c) $\omega^\star_i$ scaled over $\langle \omega_g\rangle$- the average lowest non-trivial frequencies, over samples that constitute $f_{clean}$. Note that while $\langle \gamma_i \rangle $ are monotonic functions of $r_c$, the significant reduction in $f_{clean}$ with increasing $r_c$ due to the broadening of domain in $r$ where stiffness is negative renders $r_c > 1.2$ unsuitable for fast and reliable generation of gapped samples.}
	\label{fig:spectral_mean_r_c_variation}
\end{figure}

Analogous to the analysis performed in the previous subsection that had focused on the role of density, we investigate here the dependence of gap properties on the parameter $r_c$ that controls the width of attractive interactions in the potential (\ref{eq:potential}). In particular, we consider configurations at fixed density $\rho=1$, and evaluate the probability distributions of $\gamma_i$ (Fig.\ref{fig:spectral_distribution_r_c_variation}(a,b)) and $\omega^\star_i$ (Fig.\ref{fig:spectral_distribution_r_c_variation}(c,d)) for $r_c=\{1.15,1.20,1.25\}$. Corresponding ensemble averaged quantities are shown in Fig.\ref{fig:spectral_mean_r_c_variation}.

As with the scenario encountered with increasing $\rho$ in sec. \ref{sec:rho}, the distribution profiles (Fig.\ref{fig:spectral_distribution_r_c_variation}) and mean values of $\gamma_i$ (Fig.\ref{fig:spectral_mean_r_c_variation}) shifts in the positive direction with increasing $r_c$, and this again is accompanied by a decrease in fluctuations of the characteristic frequencies at which the respective gaps occur \ref{fig:spectral_distribution_r_c_variation}(c,d). However, the limit $r_c \to \infty$ recovers qualitatively, the long range characteristics of the LJ potential that results in configurations that does not exhibit gapped spectra. 


To resolve this corundum	, we point towards the dramatic reduction in the fraction of clean samples ($f_{clean} \sim 0.1$) present in $r_c=1.25$ that we highlight is significantly lower than even its value in the highest density scenario considered in \ref{sec:rho}. This behavior stems from the broadening regime of $U(r)$ that yields negative bond stiffness with increase in $r_c$ Fig. \ref{fig:potential}(c), that are discarded when constructing the clean ensemble. These results suggest that perhaps only a very narrow region of parameter values in the vicinity of $r_c\sim 1.2$ are ideally suitable for the robust generation phononic band gaps.

Furthermore, we note that unlike the scenario encountered when varying $\rho$ the fraction of sticky and super sticky particles, do not seem to change significantly with $r_c$ even when $f_{clean}$ decreases dramatically (see inset of Fig.\ref{fig:spectral_mean_r_c_variation}(a)). Hence, direct association between the sample stickiness of individual configurations and their corresponding gap properties seem to be tenuous despite the critical nature of the attractive component in $U(r)$ in recovering linear elasticity. In fact, we do not find any meaningful correlations between sample stickiness $s$ as defined in \cite{LopezLerner2020a}, with the gap properties for all combinations of $r_c$ and $\rho$ considered. With that said, the numerical values of the Pearson coefficients ($C_{s,\gamma_i}$) indicate a weak positive correlation between sample stickiness and the gap size - 
$ 0.047 < C_{s,\gamma_1} < 0.058$ and $ 0.054 < C_{s,\gamma_2} < 0.110$, that may be of some limited use in identifying configurations with a given gap size based on its stickiness.

\subsection{System size dependence}
\label{sec:size}

\begin{figure}[htp!]
	\includegraphics[width=0.95\columnwidth]{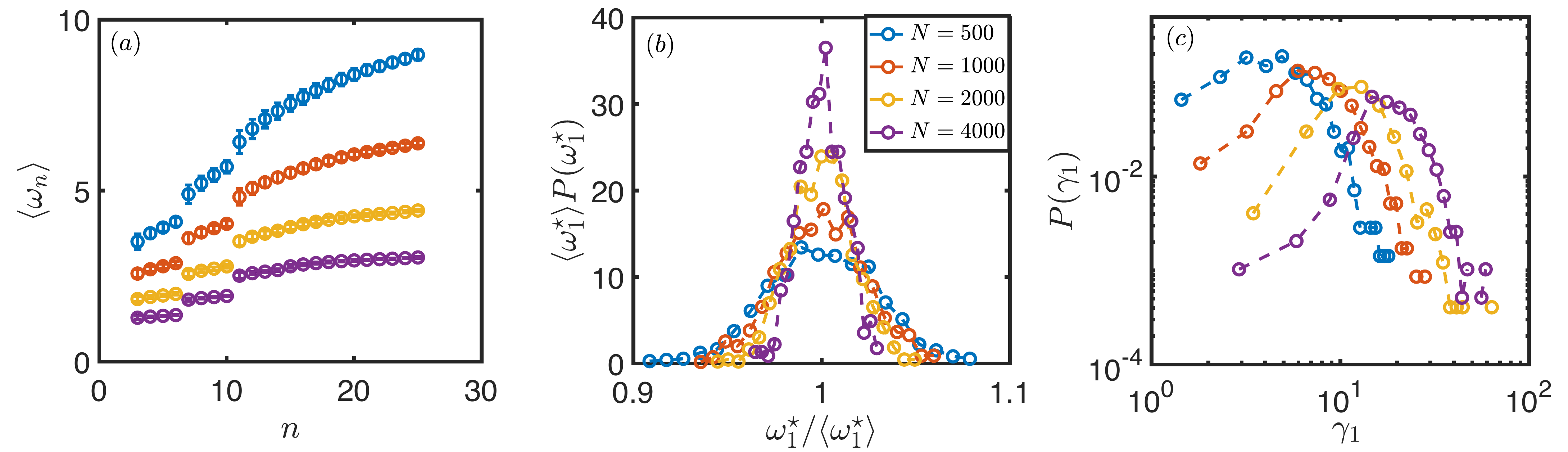}		
	\caption{(a) Average vibrational frequencies of a given mode number $n$- $\langle \omega_n \rangle$ and corresponding probability densities of (b) $\omega_1^{\star}$ and (c) $\gamma_1$ for $\rho=1.2, r_c=1.2$ for various system sizes. }
	\label{fig:size_variation}
\end{figure}

We investigate the effects of system size on the behavior and statistical properties of spectral gaps. As mentioned earlier, the predictions of linear elasticity is made increasingly precise by an increase in system size of disordered materials \cite{TanguyBarrat2002}. The use of sticky potentials in this work act to suppress the density of soft modes, such that linear elastic behavior and the occurrence of gapped spectra sets in at significantly smaller length scales. Here we show in Fig.\ref{fig:size_variation}, that these gaps persist and their statistical properties remain robust even in small systems for $\rho=1.2$ and $r_c=1.2$, (i.e parameter values that have been previously determined to be in the ideal range for gapped spectra creation).

While we see that the fluctuations in $\omega_1^{\star}$ decrease and typical values of $\gamma_1$ increase with larger system sizes as expected also for purely repulsive potentials  \cite{TanguyBarrat2002}, the spectral gap in systems as small as $N=500$ remains well-defined and pronounced. This $N$ is significantly smaller than the $N \sim 4 \times 10^5$ found by \cite{TanguyBarrat2002} needed to recover continuum elasticity in two dimensions. Lastly, we note that while the gap is unambiguously present even at $N=500$, the degeneracy of frequencies within each linear elastic branch are increasingly lifted with smaller system sizes (see Fig. \ref{fig:size_variation}(a)). This may aid in selecting or exciting particular eigenmodes for practical applications so long as the broadening of bands do not destroy the primary features of the spectral gap.

\section{Eliminating Pre-stress Contributions}	
\label{sec:pre_stress}
In this section, we discuss and show how eliminating pre-stress contributions to the dynamical matrix (\ref{eq:hessian}) can be used to obtain stress-free normal modes that we also show serve to enhance spectral gap features of the elastic network. To begin with, we write down explicitly the
off-diagonal terms $i \neq j$ of the hessian
\begin{equation}
\mathcal{M}^{\alpha \beta}_{i \neq j}= \sum_{\langle ij \rangle}  \left[ \sum_{\alpha = \beta} \left\{ -k_{ij}\frac{(R^\alpha_i - R^{\beta}_j)^2}{r^2_{ij}}+f_{ij} \left( \frac{(R^{\alpha}_i - R^{\beta}_j)^2-r_{ij}^2}{r^3_{ij}}\right)\right\} - \sum_{\alpha \neq \beta} k_{ij} l_{0,ij} \left\{\frac{(R^\alpha_i - R^{\alpha}_j)(R^{\beta}_i - R^{\beta}_j)}{r^3_{ij}} \right\} \right]
\label{eq:hessian_explicit}
\end{equation}
where $k_{ij}$ and $l_{0,ij}$ are respectively the stiffness and rest lengths of the harmonic interaction between particles $i$ and $j$ that constitute the effective elastic network. Now, in the absence of pre-stress, the terms involving $f_{ij}, l_{0,ij}$ vanishes, and we are thus able to write down a stress-free hessian $\tilde{M}$ that is completely characterized by the set of stiffnesses $\left\{ k_{ij}\right\}$:

\begin{align}
\tilde{\mathcal{M}}^{\alpha \beta}_{i \neq j}= \sum_{\langle ij \rangle}  \sum_{\alpha = \beta} -k_{ij}\frac{(R^\alpha_i - R^{\beta}_j)^2}{r^2_{ij}} && \tilde{\mathcal{M}}^{\alpha \beta}_{i = j} = -\sum_{i} \tilde{\mathcal{M}}^{\alpha \beta}_{i \neq j} 
\label{eq:stress_free}
\end{align}

Upon diagonalizing $\tilde{\mathcal{M}}$, we then obtain the vibrational spectrum of the stress-free network with corresponding eigenfrequencies $\left\{\omega'_n\right\}$ that now depends only on a set of stiffness parameters $\{k_{ij}\}$. We note that existing approaches in designing allosteric metamaterial that tune for specific functionalities, also eliminate pre-stress contributions by construction, and in this regard is unlike natural occurring proteins it is trying to simulate \cite{RocksNagel2017}. However, this reduction in the number of local elastic parameters of the network have practical advantages when it comes to the actual synthesis of the material notwithstanding its positive effects on the spectrum that we discuss subsequently. 

\begin{figure}[htp!]
	\includegraphics[width=0.95\columnwidth]{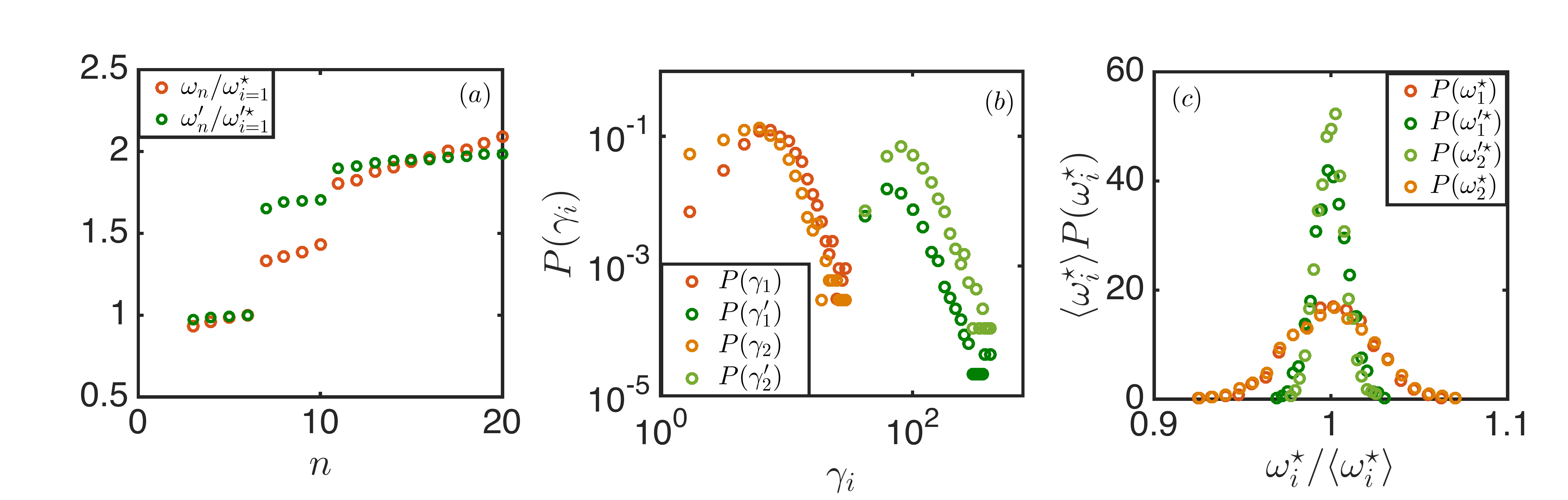}		
	\caption{(a) Vibrational frequencies $\{\omega_n, \omega'_n \}$ scaled by the gap frequency of the first branch $ \omega^\star_{i=1}, \omega'^\star_{i=1}$ against mode number ($n$) before (orange) and after (green) eliminating pre-stress contributions for a single configuration ($r_c=1.2, \rho=1$). Probability density functions of (b) $\gamma$ and (c) $\omega$ obtained from diagonalizing $\mathcal{M}$ (orange) and $\tilde{\mathcal{M}}$ (green).}
	\label{fig:pre_stress}
\end{figure}

In Fig.\ref{fig:pre_stress}(a) we plot the ordered set of eigenfrequencies for elastic networks with (orange) and without pre-stress(green) contributions derived respectively from $\mathcal{M}$ and $\tilde{\mathcal{M}}$ originating from the same energy minimal configuration of a sticky solid at $r_c=1.2$ and $\rho=1$. The frequencies are scaled by their respective gap frequency of the first linear elastic branch $\omega^\star_1$ and $\omega'^\star_1$, such that the enhancement to the first gap parameter is made visually clear. We note that this accentuation of linear elastic behavior upon elimination of pre-stress is not unique to sticky potentials, but a generic feature also present in conventional models \cite{LernerBouchbinder2018}. 

Furthermore, in Fig.\ref{fig:pre_stress}(b) we see that the probability distribution of $\gamma_i$ is shifted in the positive direction for stress-free systems indicating that the enlargement in gap size is statistically significant. Fluctuations in the corresponding gap frequencies $\omega^{\star}_i$ is also observed to decrease in Fig.\ref{fig:pre_stress}(c). These results indicate that the elimination of pre-stress contributions in the energy minimal configuration, typically enhance the gap features in the derived elastic network. Here we reiterate for clarity that unlike WCA driven systems \cite{RocksNagel2017}, these gaps are reliably present in sticky networks even without this explicit removal of pre-stress, and hence do not require any fine tuning to generate. The procedure detailed in this section serves as an optional amplification of the gaps. 

We remark that ``unclean" samples due to the appearance of bonds with negative stiffness previously discussed in section \ref{sec:rho} and \ref{sec:r_c} can be made experimentally relevant by simply discarding bonds with negative stiffness (i.e replacing all $k< 0$ by $k=0$) when considering the stress-free hessian $\tilde{\mathcal{M}}$ [\ref{eq:stress_free}]. For all densities considered at $r_c=1.2$, we find that the typical deviation of $\gamma_i$ upon removing these bonds are less than 1\% from their corresponding values in the stress-free system. 

Lastly, we note that imprecision in printing or random variability involved in experimental realization of our approach may also result in deviations in the bond stiffness $k$ used in practise. However, we find that random perturbations in the values of $k$ do not significantly affect the properties of the spectral gaps. In particular, by imposing random perturbations selected from a uniform distribution $\delta k \sim [-\bar{k}\delta_{max},+\bar{k}\delta_{max}]$, we find that the typical deviation of $\gamma_i$ upon imposing these perturbations are less than $ 3\%$ for $\delta_{max}$ as large as $0.05$.

\section{Conclusion} \label{sec:conclusion}

\begin{figure}[htp!]
	\includegraphics[width=0.7\columnwidth]{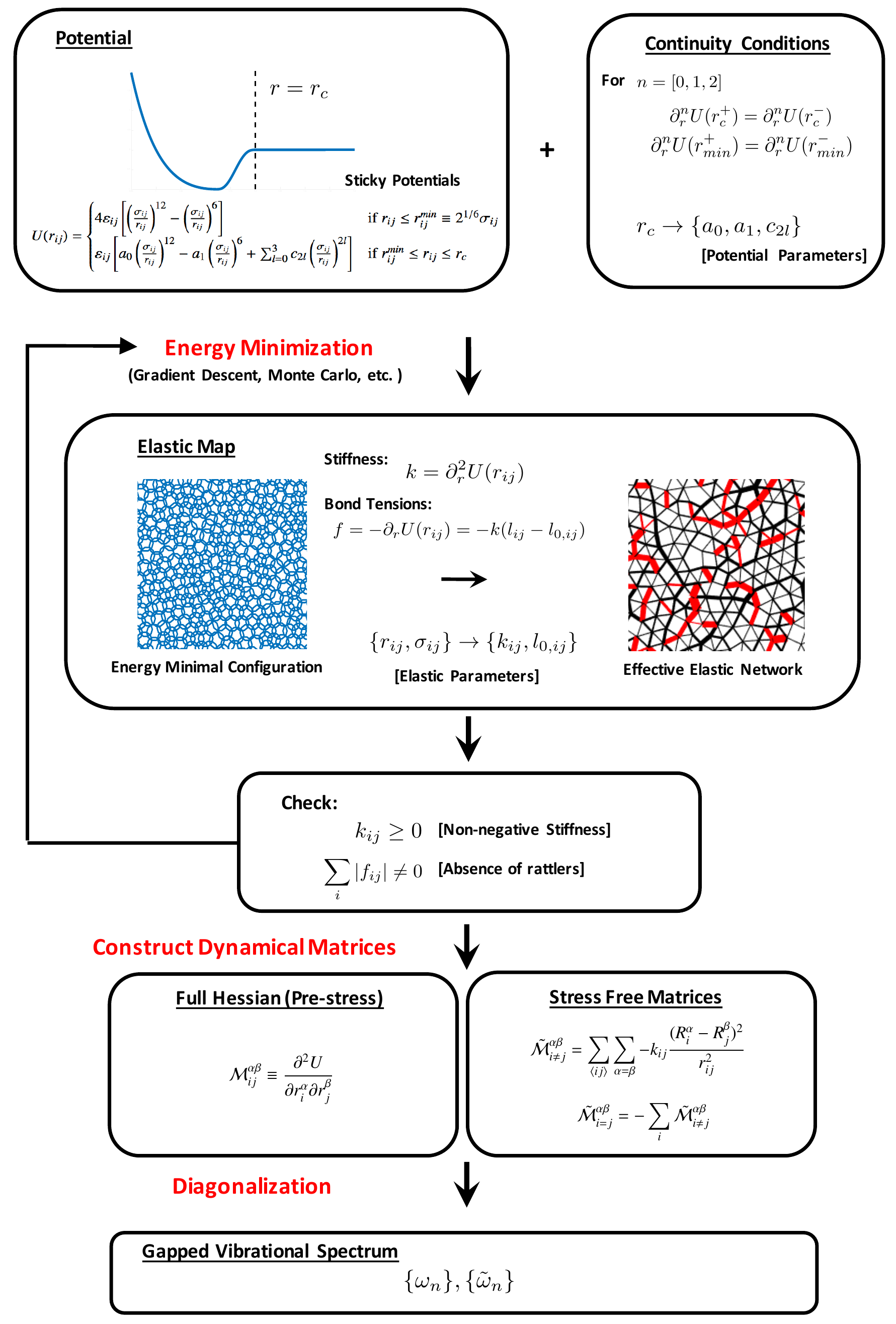}		
	\caption{Schematic illustration for fast generation of gapped phononic spectrum with no fine tunning}
	\label{fig:flow_chart}
\end{figure}

In this work, we enumerated the spectral properties of disordered solids consisting of particles mediated by a class of sticky potentials and detailed its various parameter and size dependencies. For a wide range of parameters, we show that such systems adhere to continuum elasticity even in small systems and that sizeable spectral gaps are consistently and predictably present in absence of any fine tuning. In addition, we discuss appropriate parameter selection and various strategies including the elimination of pre-stress that collectively enhance gap features for practical applications. To succinctly summarize these findings and formalize our approach, we present our results as a step-wise algorithm in generating elastic networks with robust spectral gaps that is depicted by the schematic flow chart in Fig.\ref{fig:flow_chart}.

Lastly, we emphasize that while this approach in itself, reliably generates spectral gaps without fine tuning, the adaptation of these sticky potentials in constructing elastic networks is meant to be used in conjunction with the various exisiting protocols that optimize for specific functionalities in mechanical response \cite{RocksNagel2017,RocksKatifori2019,RonellenfitschDunkel2019} or novel bulk \cite{ReidDePablo2018,ReidDePablo2019} and topological properties \cite{RocklinMao2017,SouslovVitelli2017}.  The potential enhancement of various functionalities  beyond the accentuation of spectral gaps brought about by the use of sticky potentials is thus a very interesting avenue for investigation. 
\\

\begin{acknowledgments}
We acknowledge support from the Singapore Ministry of Education through the Academic Research Fund Tier 1 	(2019-T1-001-03), Singapore and are grateful to the National Supercomputing Centre (NSCC) of Singapore for providing the computational resources.
\end{acknowledgments}
 
\end{document}